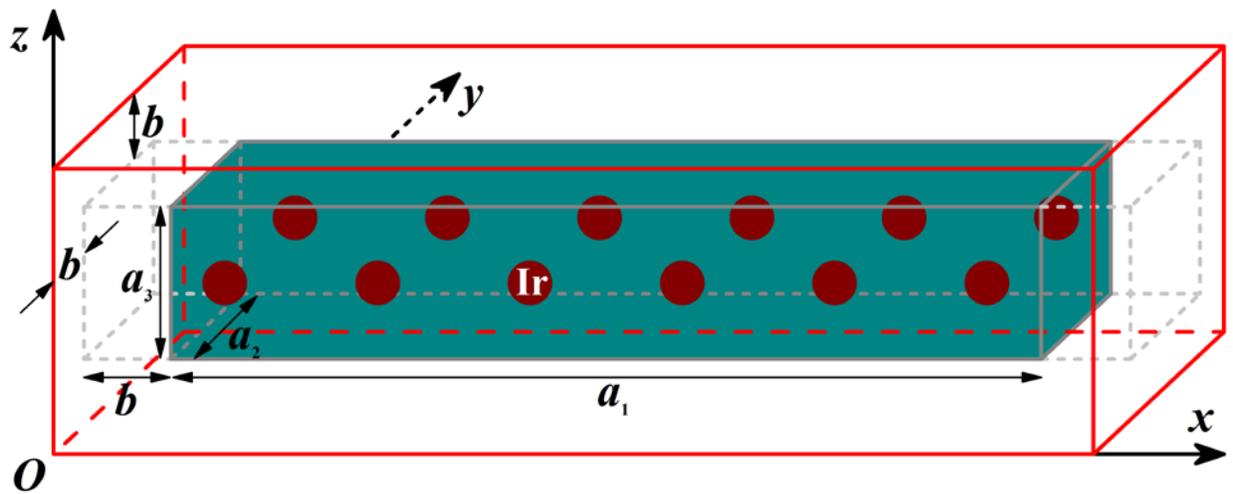

**Graphical Table of Contents**

An electron-gas model for experimentally observed self-assembled Ir nanowires grown on Ge(001) surface. These Ir nanowires have magic lengths which are an integer multiple of 4.8 nm.

# Analysis of magic lengths in growth of supported metallic nanowires

Yong Han*

Department of Physics and Astronomy, Iowa State University, Ames, Iowa 50011, USA

Ames Laboratory—U. S. Department of Energy, Iowa State University, Ames, Iowa 50011, USA

*Email: yong@ameslab.gov

**ABSTRACT**

Metallic nanowires can exhibit fascinating physical properties. These unique properties often originate primarily from the quantum confinement of free electrons in a potential well, while electron-electron interactions do not play a decisive role. A recent experimental study shows that self-assembled Ir nanowires grown on Ge(001) surface have a strong length preference: the nanowire lengths are an integer multiple of 4.8 nm. In this paper, a free electron-gas model for geometries corresponding to the nanowires is used to analyze the selection of these preferred or magic lengths. The model shows that the inclusion of even numbers of free electrons in an Ir nanowire corresponds to these magic lengths once an electron spillage effect is taken into account. The model also shows that the stability of the nanowire diminishes with its increasing length, and consequently suggests why no long nanowires are observed in experiments. It is also shown that applying generic results for quantum size effects in a nanofilm geometry is not adequate to accurately describe the length selection in the rather different nanowire geometry, where the transverse dimensions smaller than the electron Fermi wavelength. Finally, the monatomic Au chain growth on Ge(001) surface is also analyzed. In contrast to Ir nanowires, the model shows that the stability of an Au chain strongly depends on its electron spillage.

# 1. Introduction

Size-dependent features exhibited in physical properties of metallic nanoclusters, nanofilms, and nanowires are often attributed to quantum confinement effects on free electrons in a geometrically symmetric potential well [1, 2, 3, 4, 5, 6, 7, 8, 9]. This size dependence of physical properties is known as the quantum size effect (QSE). For metallic nanofilms with thickness of a few single-atom layers, the confinement of electrons by two parallel surfaces causes periodic oscillations of electronic properties as functions of film thickness [10]. Certain thicknesses of a nanofilm are particularly favorable in energy. Existence of such magic thicknesses, which can be called one-dimensional (1D) "shell" effects [6], have been verified by experiments [11, 12, 13, 14, 15, 16]. For metallic nanowires, theoretical models [17, 18, 19, 20, 21, 22] predict the two-dimensional 2D shell structures in electronic properties, i.e., the electronic properties oscillates with increasing radial dimension, and some radii of nanowires can be particularly thermodynamically stable. Indeed, these magic radii are observable in experiments [23, 24].

In contrast to the experimental observations and theoretical modeling of shell structures along radial (or lateral) directions of a metal nanowire, Mocking *et al.* have recently reported the preferred length growth (along the axial direction instead of lateral directions) of self-assembled Ir nanowires deposited on Ge(001) surface from scanning tunneling microscopy (STM) and spectroscopic measurements [25]. They found that the length distribution of Ir nanowires shows a strong preference for integer multiples of 4.8 nm, and that long wires are rare. Although they suggested quantum size interferences are responsible for these preferred lengths by comparing a generic result of QSE for nanofilm geometry, refined modeling to capture the nanowire geometry is more appropriate, as performed in this work.

Since QSE results just from the quantum confinement of free electrons in a potential well and electron-electron interactions do not play a decisive role, in this work an electron-gas model (EGM) with no electron-electron interactions is used to describe the targeted metal nanowire systems. Various choices of confining potential well are possible, but the simplest form, i.e., the infinite-depth hard-wall potential well, is sufficient for the purpose of describing the stability of this type of confinement system. However, to describe other properties, e.g., work functions, a finite-depth potential well is necessary [7]. In fact, the EGM has been very successful in description of shell structure and phase relations of both metal nanofilms [6, 7] and metal nanowires [8]. From the results of the EGM, once a certain electron spillage effect at the metal-semiconductor interface is taken into account, an *even* number ($N$ = 2, 4, 6, …) of remaining free electrons in the wire will correspond to a stable wire length. The smaller even numbers correspond to more stable lengths, while the stability of the nanowire diminishes with increasing even number of $N$ (and therefore length) because the overall surface free energy gradually increases. This feature is consistent with the fact that no long nanowires are observed experimentally [25]. From the EGM, it is also demonstrated that theoretically treating the nanowire as a nanofilm fails in the explanation of such magic growth of the nanowire, because the lateral sizes (perpendicular to the growth direction or axial direction) of the targeted Ir nanowire are too small, rather less than its corresponding electron Fermi wavelength. The oscillation behavior of electronic properties of an ultra thin metal nanowire with increasing length is completely different from those of a metal nanofilm with increasing thickness, as will be illustrated by plotting the curves of Fermi energy (or chemical potential), surface free energy, and electron density versus wire length in following sections.

Finally, for comparison with Ir nanowires, the EGM is also used to describe a straight monatomic Au chain grown on Ge(001) surface. It is found that an isolated (i.e., unsupported in vacuum) monatomic Au chain can have the similar stability behavior to the Ir nanowires on Ge(001) surface, but it will no longer show the preferred lengths after a certain amount of electron spillage is reached.

## 2. Models, results, and discussion

### 2.1 Ir nanowires on Ge(001) surface

The STM images from the experiments [25] shows that the arrangement of atoms of Ir nanowires adopts a well-ordered double chain geometry running perpendicular to the direction corresponding to that of the Ge dimer rows on the original, clean (reconstructed) Ge(001) terrace. The experimentally observed Ir nanowires have a constant width of 0.57 nm. The periodicity along the growth direction of a wire is 0.8 nm, i.e., the double of Ge(001) surface lattice constant (0.4 nm). For more details, see Figs. 1(a) – 1(d) of Ref. [25].

To mimic the geometry of the Ir nanowires, a reasonable choice of the potential-well geometry is a rectangular box, as illustrated in Fig. 1. As we discussed previously [7, 8], the charge neutrality requires a dependency of the potential-well-boundary position on the corresponding barrier potential-well height $U_0$ in a free-electron-gas model for a metal surface. For $U_0 \rightarrow \infty$, i.e., the hard-wall potential well, the separation, $b$, between the potential-well boundary surface and geometric surface of uniform positive-charge background satisfies $b = 3\lambda_F/16$, where $\lambda_F$ is the electron Fermi wavelength for a bulk metal in the standard free-electron-gas model, i.e., the Drude-Sommerfeld model, and solely determined by the bulk average electron density

$$w_0 \equiv \frac{N}{V} = \frac{8\pi}{3\lambda_F^3}. \tag{1}$$

Here $N$ is the number of electrons contained in a bulk volume $V$. Simply setting $b$ to zero can lead to a significant discrepancy between the results from the model and density functional theory (DFT) calculations [7]. Therefore, $b > 0$ has to be considered when the EGM is used, as depicted in Fig. 1. In this work, $b = 3\lambda_F/16$ is taken.

For conciseness of notation, if $a_1$, $a_2$, and $a_3$ measure the geometric dimensions of the rectangular box along $x$, $y$, and $z$ directions, respectively, the potential-well dimensions are defined as $A_1 \equiv a_1 + 2b$, $A_2 \equiv a_2 + 2b$, and $A_3 \equiv a_3 + 2b$ (see Fig. 1). If the Cartesian coordinate system in Fig. 1 is chosen, the eigenenergies of the Schrödinger equation for an electron in the hard-wall potential well (the box with red lined edges in Fig. 1) can be expressed as

$$\epsilon_{n_1,n_2,n_3} = \frac{\pi^2\hbar^2}{2m_e}\left(\frac{n_1^2}{A_1^2} + \frac{n_2^2}{A_2^2} + \frac{n_3^2}{A_3^2}\right), \tag{2}$$

where each of quantum numbers $n_1$, $n_2$, and $n_3$ takes all positive integers, and $m_e$ is the rest mass of an electron. The corresponding wavefunctions of an electron are expressed as

$$|n_1, n_2, n_3\rangle \equiv \psi_{n_1,n_2,n_3}(x,y,z) = \sqrt{\frac{8}{A_1 A_2 A_3}} \sin\frac{n_1 \pi x}{A_1} \sin\frac{n_2 \pi y}{A_2} \sin\frac{n_3 \pi z}{A_3}. \tag{3}$$

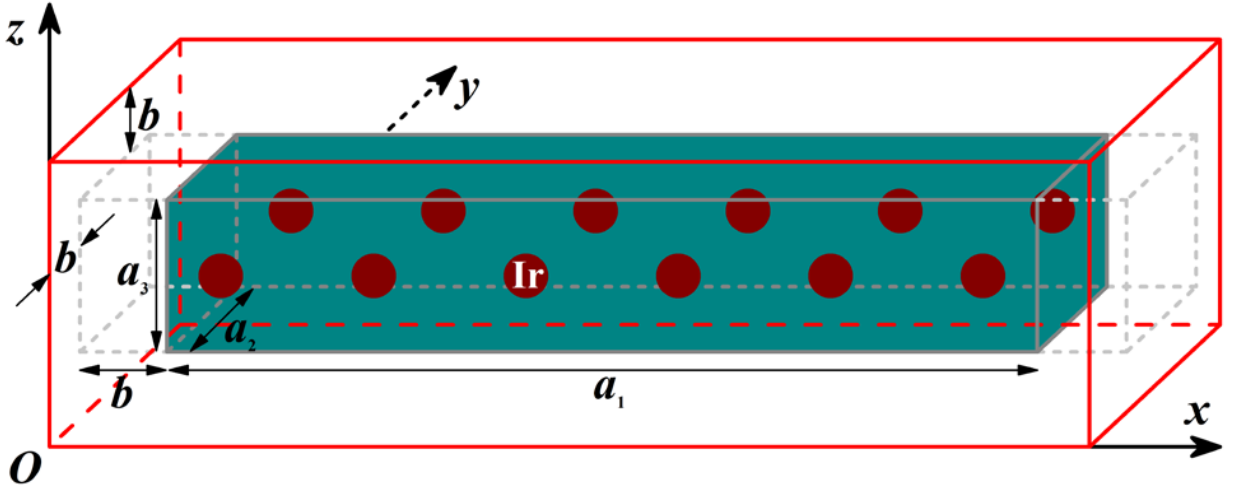

**Figure 1** Schematic diagram illustrating an Ir nanowire in a rectangular-box potential well with the surface charge-neutrality requirement. The shadowed box with gray lined edges represents the geometry of uniform positive-charge background, the box with red lined edges represents the potential-well boundaries, and the spheres represent the positions of ionic cores of Ir atoms.

To assess the stability of a nanowire during growth, one needs to calculate the surface free energy, which can be defined as

$$s = \frac{E - N\sigma_0}{2(a_1 a_2 + a_2 a_3 + a_3 a_1)}, \tag{4}$$

where $N$ is the number of free electrons in potential well, and $E$ is the total energy of all free electrons, and $\sigma_0 \equiv \frac{E}{N} = \frac{3}{5} E_F$ is the energy per electron in the corresponding bulk metal with the average electron density $w_0$ in Eq. (1). The bulk Fermi energy,

$$E_F = \frac{\hbar^2}{2m_e} \frac{(2\pi)^2}{\lambda_F^2}, \tag{5}$$

is solely determined by $\lambda_F$, and thus by $w_0$. The total energy $E$ can be obtained from

$$E = \sum_{\text{occ.}} \epsilon_{n_1, n_2, n_3}, \tag{6}$$

where the summation is over all occupied eigenstates $|n_1, n_2, n_3\rangle$ with twofold spin degeneracy of electrons. The Fermi energy, $\epsilon_F$, or chemical potential, of the nanowire is defined as the highest energy level occupied by an electron at the ground state. The electron density at any specific point $(x, y, z)$ can be calculated by

$$w(x, y, z) = \sum_{\text{occ.}} \psi^*_{n_1,n_2,n_3}(x, y, z) \psi_{n_1,n_2,n_3}(x, y, z), \tag{7}$$

where $\psi^*_{n_1,n_2,n_3}$ is the complex conjugate of $\psi_{n_1,n_2,n_3}$.

It can be verified that the energies expressed in Eqs. (2) and (6) are dimensionless when they are in unit of $E_F$. Likewise, both $s/s_0$ and $w/w_0$ are also dimensionless, where $s_0 = \pi E_F / (20\lambda_F^2)$ is the corresponding surface free energy for bulk film [7], and bulk average electron density $w_0$ satisfies Eq. (1). Therefore, the curve shapes of these quantities versus the geometric sizes are solely determined by $\lambda_F$ or $w_0$ for a chosen potential well, and then generally apply to any species of metal with a specific $\lambda_F$ or $w_0$.

Because the Ir nanowires are supported on Ge(001) surface, a certain number of electrons of Ir atoms spill across the metal-semiconductor interface. When equilibrium is reached, the Fermi energy of the metal lines up with the charge neutrality level in the gap of the semiconductor [5]. The electron spillage leaves a dipole layer at the interface, which plays a role of electron screening [3, 26]. Due to the screening, the number, $N$, of remaining free electrons in the supported nanowire will be conserved, and therefore the nanowire can be treated as a canonical system with $N$ electrons [8]. In addition, strictly speaking, because the bottom face of the rectangular box in Fig. 1 touches the interface, the value of the parameter $b$ for the bottom face should be somehow different from that for other free faces. For simplicity, such difference is ignored in the EGM.

In principle, the amount of electron spillage can be roughly estimated from the intrinsic physical parameters of substrate and supported metal [5]. However, such estimation has some uncertainty. Furthermore, Ir is a transition-metal element with the valence-electron configuration of $5d^7 6s^2$, and only few of these valence electrons should be free. Therefore it is difficult to theoretically set an unambiguous value of the number of free electrons per unit cell (i.e., the average electron density $w_0$). In this work, the number of free electrons per unit cell in the Ir nanowire can be ultimately determined by comparing the results from the model with those from experiment, as described below.

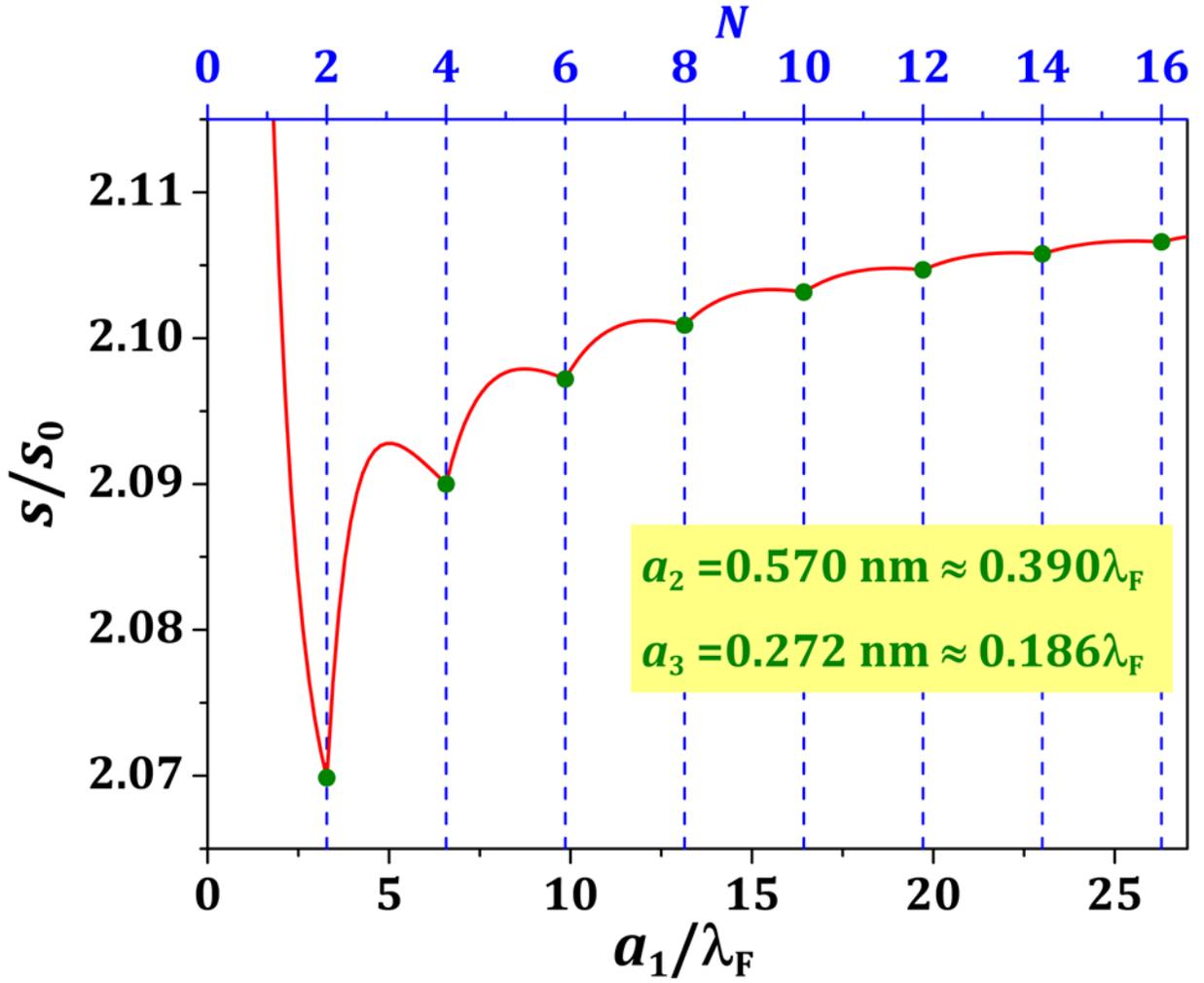

**Figure 2** Surface free energy $s$ (in unit of $s_0$) from Eq. (4) versus nanowire length $a_1$ (in unit of $\lambda_F$). Dashed blue vertical lines correspond to the stable wire lengths with even number $N$ of free electrons.

Figure 2 shows the EGM results of surface free energy $s$ as a function of nanowire length $a_1$. In the calculations, $a_2$ is set to be the experimentally observed value 0.57 nm, and $a_3$ is taken to be 0.272 nm, i.e., the ionic diameter of an Ir atom [27], because the $z$ dimension of the nanowire is of one-atomic-layer thickness. From Fig. 2, it is immediately seen that the curve of $s$ versus $a_1$ is odd-even oscillatory with increasing number $N$ of free electrons, and the minima of $s$ always correspond to even $N$ for a given value of $\lambda_F$ which is not too small. However, the $a_1$ values corresponding to these minima are varied for different $\lambda_F$ or $w_0$. The model results show that once $\lambda_F$ is taken to be ~1.461 nm (and then $a_2/\lambda_F \approx 0.390$, $a_3/\lambda_F \approx 0.186$, and $w_0 = \frac{8\pi}{3\lambda_F^3} \approx 2.687$ nm$^{-3}$), the $a_1$ values corresponding to these minima will be 4.8, 9.6, 14.4, … nm (i.e., the magic lengths observed in experiment), as seen in Fig. 2. According to Ref. [25], an experimentally extracted value of 1D Fermi wavelength is 9.6 nm, which can be shown to be equivalent to the above 3D Fermi wavelength value of ~1.461 nm. Thus, the model result in Fig. 2 is in perfect agreement with experimental observations.

Because the periodicity along the growth direction of an Ir nanowire is twice the Ge(001) surface lattice constant (0.4 nm), the smallest magic length 4.8 nm is equivalent to a distance of six unit cells with the length 0.8 nm. As discussed above, the number of free electrons is $N = 2$ within this length, and then the average electron density is 1/3 per unit cell (containing two Ir atoms), i.e., each Ir atom contributes 1/6 free electrons. Here it is should be mentioned that, to get a smooth surface free energy curve (e.g., the red curve in Fig. 2), the continuity of free-electron charge has been assumed in the calculations. If Fermi energy level is occupied by a fraction charge between 0 and 2 electrons, its energy contribution is calculated as the product of the Fermi energy and the fraction, which can be understood as the "occupation probability". Obviously, the surface free energy curve in Fig. 2 is completely different from that for nanofilms in Ref. [7]. More discussion details will be given below about this.

To reveal why the odd-even oscillations occur, let us examine how the free electrons occupy quantum states. When the lateral dimensions $a_2$ and $a_3$ are rather less than $\lambda_F$ (e.g., the Ir nanowires discussed above), the quantum numbers $n_2$ and $n_3$ of occupied states are actually always the constant integer of 1. As $a_1$ increases, electrons only occupy the quantum states with $n_1$ taking larger integers, i.e., the 3D potential well is reduced to be 1D except for a constant term related to $n_2 = 1$ and $n_3 = 1$. With increasing $N$, the electrons successively occupy the twofold degenerate state $|1,1,1\rangle$ for $N =1$ and then 2, state $|2,1,1\rangle$ for $N =3$ and then 4, state $|3,1,1\rangle$ for $N =5$ and then 6, and so on. Thus, the twofold degeneracy due to electron spin gives rise to the odd-even oscillation behavior. Consequently, the occupied highest energy level $\epsilon_F$ (i.e., Fermi energy or chemical potential) is odd-even oscillatory with increasing $a_1$ and therefore $N$, as shown in Fig. 3(a).

If the lateral dimensions $a_2$ and $a_3$ are close to or larger than $\lambda_F$, the quantum numbers $n_2$ and $n_3$ of occupied states will be not constant, and take different integers greater than 1. This results in irregular curves of $\epsilon_F$ versus $N$. See Figs. 3(b) and 3(c). As $a_2$ and $a_3$ are taken to be larger and larger, the plot approaches a smooth curve with cusps (see Fig. 3(d)), and $\epsilon_F$ oscillates around $E_F$ with an approximate period of $\lambda_F/2$. i.e., the nanofilm limit ($a_2$ and $a_3$ are infinite and $a_1$ is understood as the thickness of a nanofilm), which has already been discussed in details in Ref. [7].

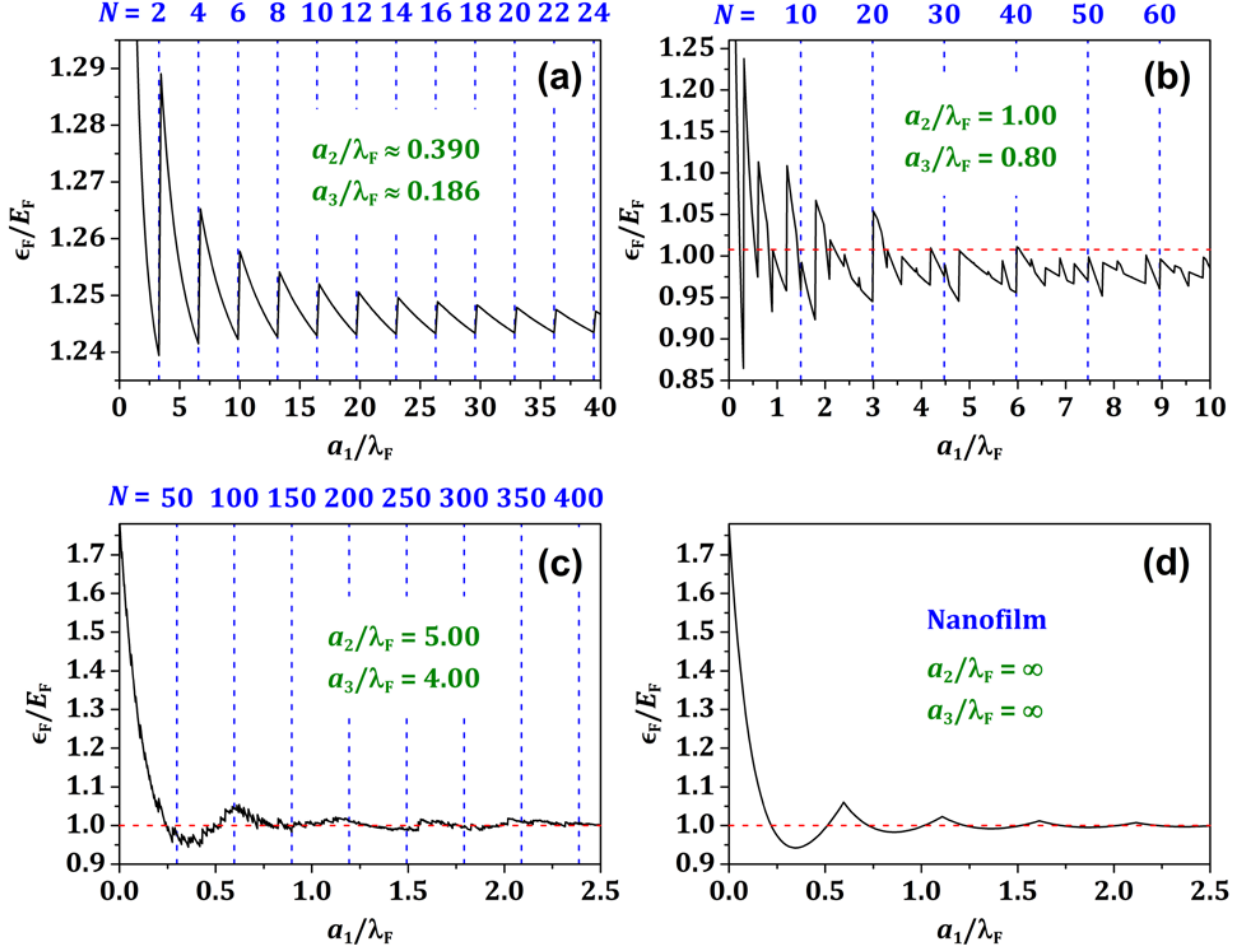

**Figure 3** Fermi energy $\epsilon_F$ from Eq. (2) versus axial dimension $a_1$ with different lateral dimensions $a_2$ and $a_3$: (a) $a_2/\lambda_F \approx 0.390$, $a_3/\lambda_F \approx 0.186$; (b) $a_2/\lambda_F = 1.00$, $a_3/\lambda_F = 0.80$; (c) $a_2/\lambda_F = 5.00$, $a_3/\lambda_F = 4.00$; (d) $a_2/\lambda_F = \infty$, $a_3/\lambda_F = \infty$. Dashed blue vertical lines correspond to the number of free electrons in the rectangular-box potential well. Dashed red horizontal lines indicate $\epsilon_F = E_F$.

Figure 4 shows a series of curves of surface free energy $s$ versus axial dimension $a_1$ with different lateral dimensions $a_2$ and $a_3$. Relative to $\lambda_F$, if $a_2$ and $a_3$ are too small, e.g., $a_2/\lambda_F = 0.25$ and $a_3/\lambda_F = 0.20$, see Fig. 4(a), only $N = 2$ corresponds to a minimum and then stable length. Other even $N$ correspond to concave points but not minima, and then much less stable lengths. When $a_2$ and $a_3$ are rather bigger, e.g., $a_2/\lambda_F = 0.50$ and $a_3/\lambda_F = 0.40$, see Fig. 4(b), all even numbers of $N$ correspond to minima and therefore stable lengths, but larger even $N$ or longer lengths have lower $s$, i.e., longer nanowires are more stable. This is not consistent with experiments, where longer nanowires were not more favorable. As $a_2$ and $a_3$ are close to $\lambda_F$, the curve becomes irregular and nonperiodic. With increasing $a_1$, the overall curve of $s$ gradually approaches the bulk film value $s_0$, but still always higher than $s_0$, as indicated by the dashed red line in Fig. 4(c). If $a_2/\lambda_F$ and $a_3/\lambda_F$ are taken to be further bigger, see Figs. 4(d) and 4(e), the plot of $s$ tends towards a continuous decaying sinusoidal-like curve with an approximate period of $\lambda_F/2$, i.e., the nanofilm limit [7], where $s$ oscillates around $s_0$, see Fig. (f).

At first glance, it seems that the preferred magic lengths of Ir nanowires observed in experiment may correspond to the minima of $s$ curve from nanofilm limit in Fig. (f). If this is true, then the first minimum position of $\sim \lambda_F/2$ on the $s$ curve should be approximately equal to the first magic length 4.8 nm, i.e., $\lambda_F = \sim 9.6$ nm. According to Eq. (1), each unit cell only contributes to $\sim 0.00117$ free electrons, i.e., an Ir nanowire of 4.8 nm contains $\sim 0.007$ electrons. This is obviously an unphysical result, and then using the nanofilm limit cannot explain the magic lengths. Thus, physically reasonable curve corresponding to the magic lengths of Ir nanowires should be that shown in Fig. 2.

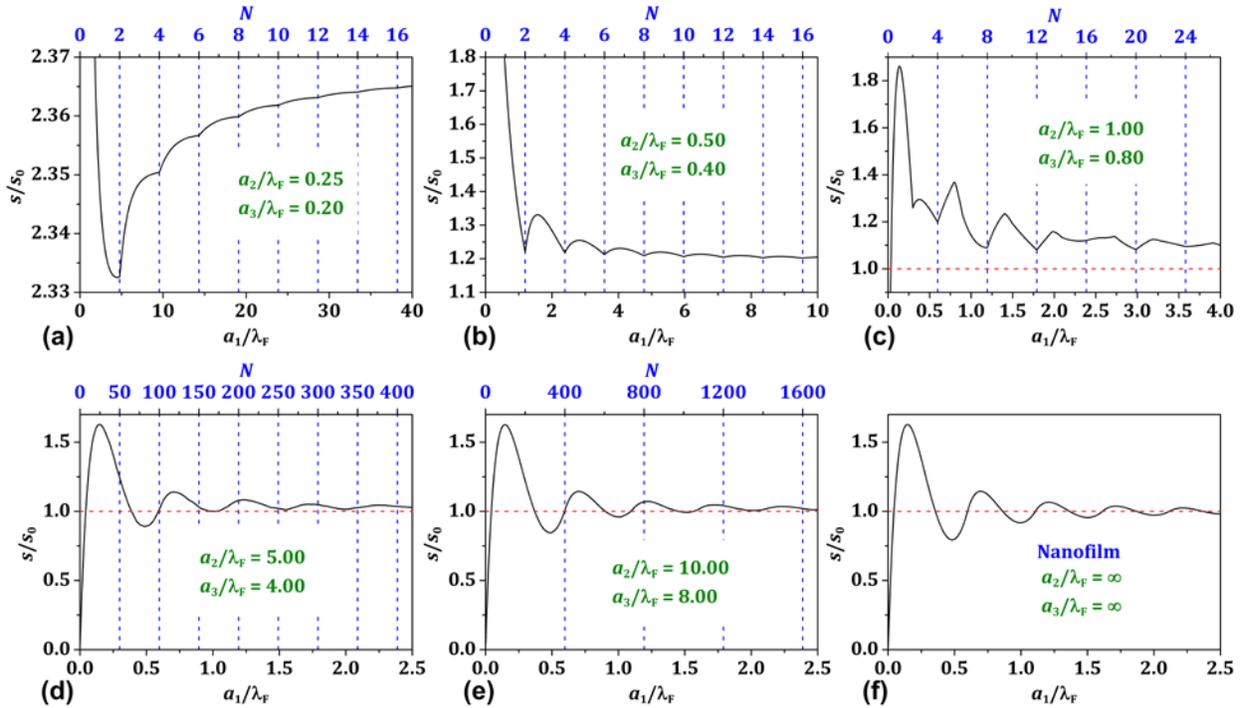

**Figure 4** Surface free energy $s$ (in unit of $s_0$) from Eq. (4) versus axial dimension $a_1$ with different lateral dimensions $a_2$ and $a_3$: (a) $a_2/\lambda_F = 0.25$, $a_3/\lambda_F = 0.20$; (b) $a_2/\lambda_F = 0.50$, $a_3/\lambda_F = 0.40$; (c) $a_2/\lambda_F = 1.00$, $a_3/\lambda_F = 0.80$; (d) $a_2/\lambda_F = 5.00$, $a_3/\lambda_F = 4.00$; (e) $a_2/\lambda_F = 10.00$, $a_3/\lambda_F = 8.00$; (f) $a_2/\lambda_F = \infty$, $a_3/\lambda_F = \infty$. Dashed blue vertical lines correspond to the number of free electrons in the rectangular-box potential well. Dashed red horizontal lines indicate $s = s_0$.

The spatial electron densities of a nanowire are closely related to its conductance, and can be calculated from Eq. (7). Figure 5 selectively shows the ground-state electron densities at the end and middle centers of the geometric rectangular box with increasing axial dimension $a_1$ for different lateral dimensions $a_2$ and $a_3$. Again, one can see that the shapes of electron density curves for the lateral dimensions rather less than $\lambda_F$ are very different that for nanofilm limit. The end center electron density $w_{end}$ is overall smaller than the middle center electron density $w_{mid}$ due to the Bardeen-Friedel oscillations, i.e., the electron density oscillates from middle center to end surface and then decays across the end surface [28, 29, 7]. As $a_2$ and $a_3$ become bigger, $w_{end}/w_0$ approaches a constant $\kappa = 0.4557 \ldots$ with a period of $\sim \lambda_F/2$, and $w_{mid}/w_0$ approaches 1 with a period of $\sim \lambda_F$, as shown in Fig. 5.

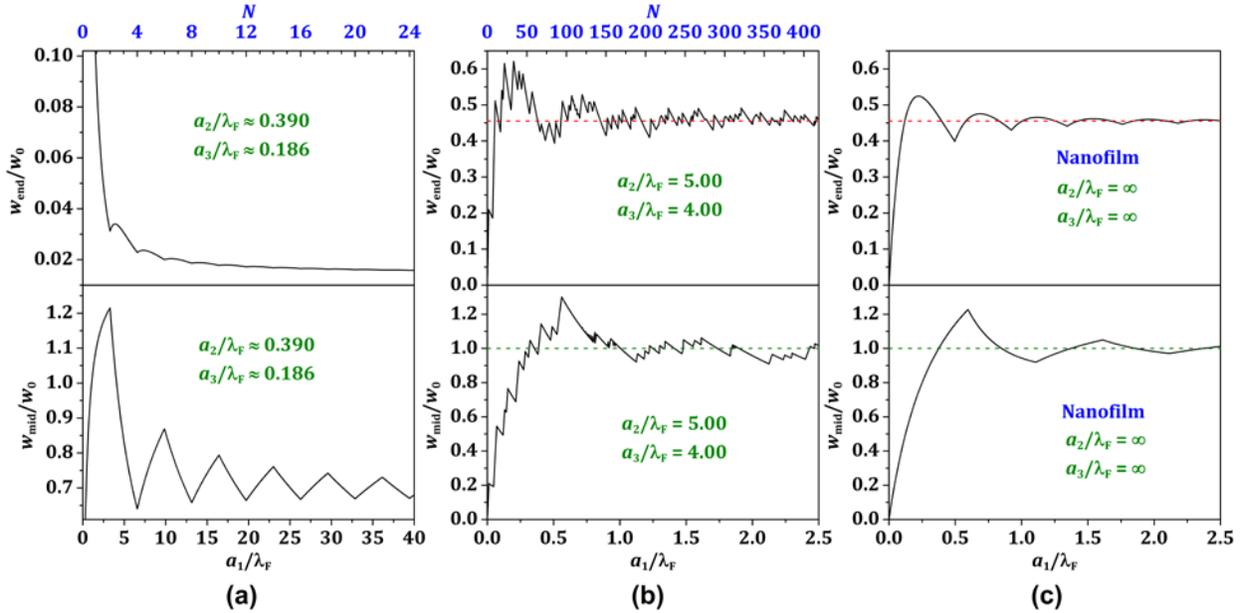

**Figure 5** Electron densities from Eq. (7) versus axial dimension $a_1$ with different lateral dimensions $a_2$ and $a_3$: (a) $a_2/\lambda_F \approx 0.390$, $a_3/\lambda_F \approx 0.186$; (b) $a_2/\lambda_F = 0.50$, $a_3/\lambda_F = 0.40$; (c) $a_2/\lambda_F = \infty$, $a_3/\lambda_F = \infty$. $N$ corresponds to the number of free electrons in the rectangular-box potential well. $w_{end}$ is the density at the end center ($x = b + a_2/2, y = b + a_3/2, z = b$) of the geometric rectangular box, and $w_{mid}$ is the density at the middle center ($x = b + a_2/2, y = b + a_3/2, z = b + a_1/2$). Dashed red horizontal lines denote the constant $\kappa = 0.4557 \ldots$ [7]. Dashed green horizontal lines indicate $w = w_0$.

## 2.2 Monatomic Au chains on Ge(001) surface

In 2004, Wang *et al*. [30] deposited Au atoms onto Ge(001), leading to self-organized Au monatomic chains on reconstructed Ge(001) surfaces. These chains can extend over several hundred nanometers in length. Similar results were reported by other groups for a range of similar experimental conditions [31, 32, 33, 34, 35, 36]. The distance between two Au atoms in a chain is 0.8 nm, i.e., twice of the Ge(001) surface lattice constant, but unlike the Ir nanowires discussed above, these observed Au chains are not isolated on the surface, there being a separation of 1.6 nm between two nearest neighbors, i.e., these Au chains form chain arrays [30, 31]. This plausibly indicates that an isolated monatomic Au chain is not favorable on the Ge(001) surface, as analyzed below.

A metal monatomic chain is cylindrically symmetric, and then choosing a cylindrical geometry to model the chain is more appropriate, and two transverse dimensions $a_1$ and $a_2$ of a rectangular box are reduced to a single dimension (wire radius $r$), as illustrated in Fig. 6. Similar to Sec. 2.1, the surface charge neutrality requires the potential-well radius $R = r + 2b_r$, where $r$ is the radius of the cylindrical uniform positive-charge background ; potential-well length $L = l + 2b_l$, where $l$ is the length of the positive-charge background. The separation $b_r$ should be slightly different from $b_l$, but both of them are taken to be $3\lambda_F/16$ in the following calculations, the conclusion being not significantly influenced by small variations of them.

After choosing the cylindrical coordinates $(\rho, \varphi, z)$ in Fig. 6, the wave functions of a single electron can be expressed as [37]

$$|m,\alpha,n\rangle \equiv \psi_{m,\alpha,n}(\rho,\varphi,z) = \frac{\sqrt{2}}{R\sqrt{\pi L}J_{m+1}(\zeta_{m,\alpha})}J_m\left(\frac{\zeta_{m,\alpha}}{R}\rho\right)e^{im\varphi}\sin\frac{n\pi z}{L}, \qquad (8)$$

where quantum numbers $m = 0, \pm 1, \pm 2, \pm 3, ...$, $\alpha = 1, 2, 3, ...$ ($\alpha \neq 1$, if $m \neq 0$), and $n = 1, 2, 3, ...$; $\zeta_{m,\alpha}$ is the αth zero of the Bessel function $J_m(x)$ with order $m$. The eigenenergies are

$$\epsilon_{m,\alpha,n} = \frac{\hbar^2}{2m_e}\left(\frac{\zeta_{m,\alpha}^2}{R^2} + \frac{n^2\pi^2}{L^2}\right). \qquad (9)$$

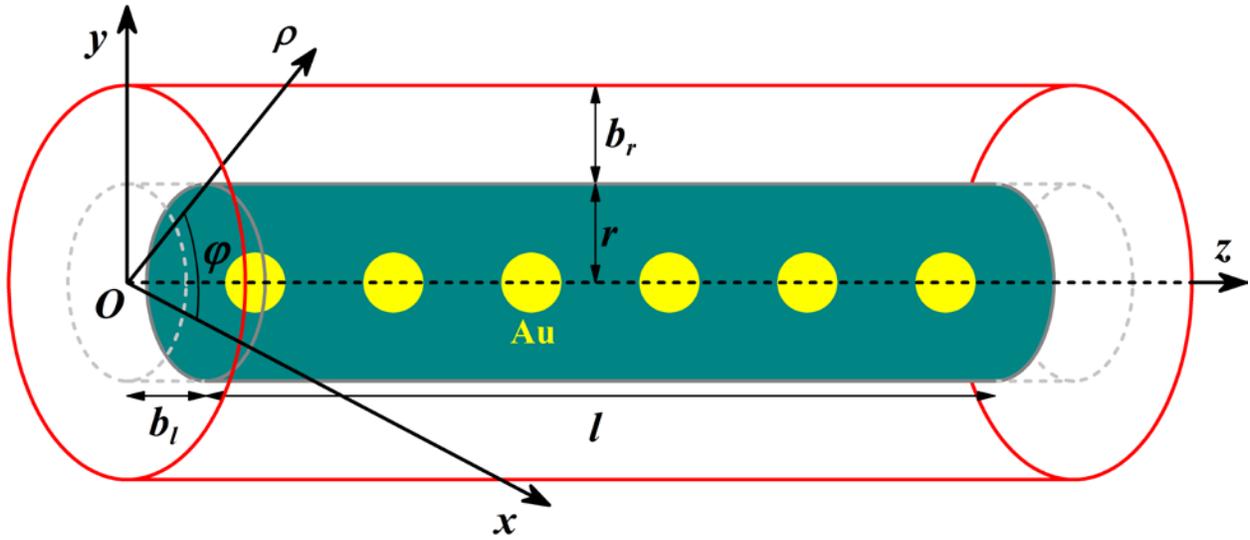

**Figure 6** Schematic diagram illustrating an Au monatomic chain in a cylindrical potential well with the surface charge-neutrality requirement. The shadowed cylinder with solid gray edges represents the geometry of uniform positive-charge background, the cylinder with solid red edges represents the potential-well boundaries, and the spheres represent the positions of ionic cores of Au atoms.

The surface free energy can be calculated as

$$s = \frac{E - N\sigma_0}{2\pi r(r+l)}, \quad (10)$$

where $N$ is the number of free electrons in potential well. The total energy of all free electrons can be calculated by

$$E = \sum_{\text{occ.}} \epsilon_{m,\alpha,n}, \quad (11)$$

where the summation is over all occupied eigenstates $|m, \alpha, n\rangle$. Note that the states $|m, \alpha, n\rangle$ and $|-m, \alpha, n\rangle$ correspond to the same eigenenergy. After taking into account the spin degeneracy of each electron state, $|\pm m \neq 0, \alpha, n\rangle$ is fourfold degenerate, and $|m = 0, \alpha, n\rangle$ is twofold degenerate. The ground-state electron density at point $(\rho, \varphi, z)$ can be obtained from the expression

$$w(\rho, \varphi, z) = \sum_{\text{occ.}} \psi^*_{m,\alpha,n}(\rho, \varphi, z) \psi_{n_1,n_2,n_3}(\rho, \varphi, z), \quad (12)$$

where $\psi^*_{m,\alpha,n}$ is the complex conjugate of $\psi_{m,\alpha,n}$.

Au atom has the valence-electron configuration of $5d^{10}6s^1$, and the 6s electron is very weakly bonded by its inert ionic core. For an isolated Au chain in vacuum with a typical atom-atom spacing (e.g., twice of the Ge(001) surface lattice constant), all s electrons can be considered to be free. If an Au chain is supported on a surface, a certain number of electrons in Au atoms spill across the interface, as analogously discussed in Sec. 2.1. The amount of electron spillage strongly depends on the intrinsic properties of the interface. For the Au chains supported on Ge(001) surface, it is difficult to exactly obtain the amount of electron spillage. Then, one can define a quantity β, which is the number density of remaining free electrons per unit cell after spilling of electrons, as an adjustable parameter. For the Au chain, the length of unit cell is 0.8 nm, i.e., twice of Ge(001) surface lattice constant (0.4 nm). It follows that $\beta = 1$ means an isolated Au chain with no electron spillage, while $\beta = 1/3$ means 2/3 of electrons spill across the interface, and so on. The radius $r$ of a monatomic Au chain is fixed to be the value of 1.44 nm from ionic radius of Au atom [27].

Figure 7 shows a series of curves of surface free energy $s$ from Eq. (10) versus length $l$ for an Au monatomic chain with different β values. Comparing Fig. 7(a) with Fig. 2, the curves are very similar, indicating that an isolated Au chain in vacuum has very similar stability to the Ir nanowires on Ge(001) surface. The magic lengths of the isolated chain are a multiple of 1.6 nm, corresponding to even $N = 2, 4, 6, …$ Decreasing β to 2/3 does not significantly change the oscillation behavior of stability, see Fig. 7(b). Further decreasing β to 1/2 results in less stable lengths for even numbers of $N > 2$, see Figs. 7(c) and 7(d), because there are no minima beyond $N = 2$. As β continues to decrease, even the length corresponding to $N = 2$ also becomes no longer stable due to the disappeared minimum of $s$ curve, see Figs. 7(e) and 7(f). Therefore, when the electron spillage reaches a certain amount, no magic lengths exist.

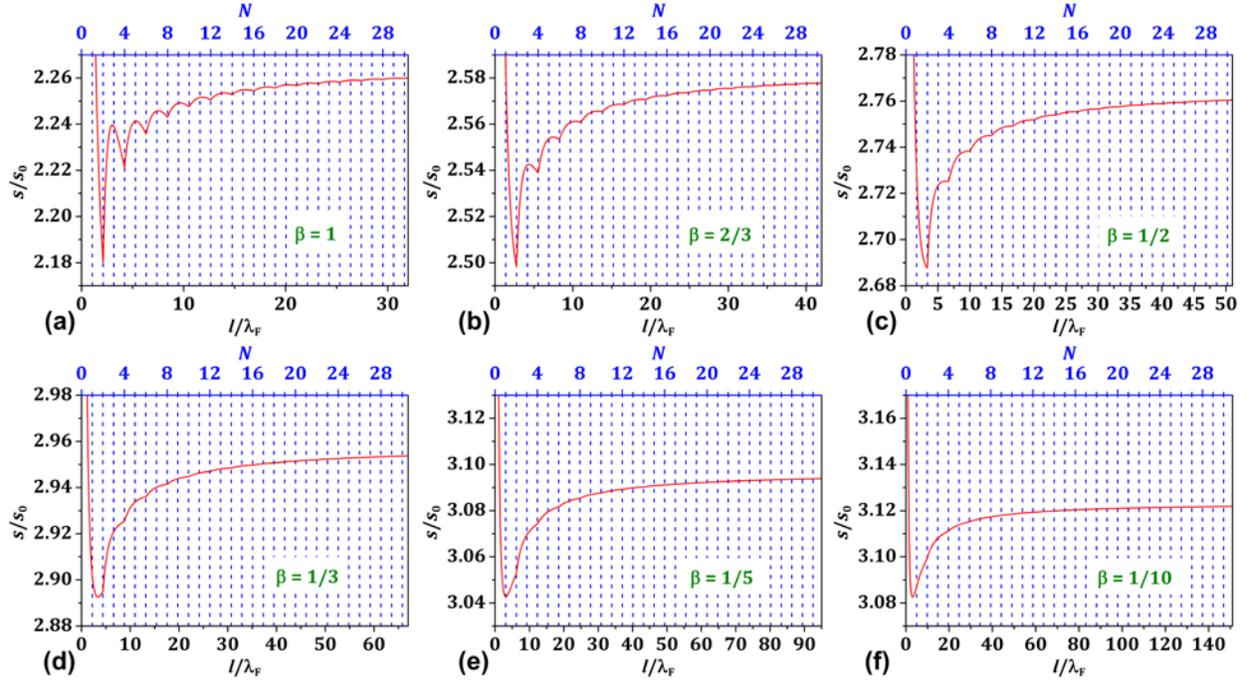

**Figure 7** Surface free energy $s$ (in unit of $s_0$) from Eq. (10) versus length $l$ (in unit of $\lambda_F$) for an Au monatomic chain. β is the number density of free electrons per unit cell. (a) $\beta = 1$; (b) $\beta = 2/3$; (c) $\beta = 1/2$; (d) $\beta = 1/3$; (e) $\beta = 1/5$; (f) $\beta = 1/10$. Dashed blue vertical lines correspond to the number of free electrons in the potential well.

In the experiments of Au deposition on Ge(001) surface, no preferred lengths of Au chains are observed, indicating electron spillage is approximately more than half. In fact, as shown in Fig. 7, if the electron spillage is large enough, a monatomic Au chain is not favorable because the $s$ value monotonically increases, and then it is not possible to obtain a stable isolated Au chain on Ge(001) surface. However, other arrangements of Au atoms are possible, as observed in experiments, e.g., Au chain arrays [30, 31] rather than isolated Au chains. DFT calculations also suggest that both simple linear and zigzag Au chains on Ge(001) surface are energetically unfavorable [38], being consistent with the above model prediction.

## 3. Concluding Remarks

In conclusion, the EGM can be used to analyze the stability of a supported metallic nanowire or a monatomic chain, and the key physics can be effectively captured. When the lateral dimensions of a nanowire are small enough relative to the corresponding electron Fermi wavelength, the 3D potential well describing electron confinement behaves as 1D, and then free electrons alternatively occupy only axial quantum states but not lateral quantum states. The electron-spin degeneracy of quantum states results in the oscillations of energy and electron densities with increasing axial dimension size. As a result, even numbers ($N = 2, 4, 6, ...$) of free electrons in an Ir nanowire corresponds to the preferred magic lengths 4.8 nm, 9.6 nm, 14.4 nm, ... if the electron spillage effect due to the metal-semiconductor interface is included. The EGM results also show that the stability of the nanowire diminishes with its increasing length because of the overall increasing of surface free energy, and consequently explains the experimental observation that no long Ir nanowires grown on Ge(001) surface are found from STM images. In contrast to Ir nanowires, a monatomic Au chain grown on Ge(001) surface does not show preferred magic lengths if the electron spillage is approximately more than half.

The EGM can generally apply to other species of metallic nanowires or other nanostructures with a specific geometry. For example, in Appendix, the EGM is also used to analyze the stability of magic radii of single-atomic-layer Ag nanopucks grown on Pb(111) surface, showing the results which are in reasonable agreement with those from experiments [39] and DFT calculations [39, 40]. Oncel *et al.* [41] used a 1D free-electron model, which is similar to the EGM, to analyze the self-organized Pt-nanowires arrays on Ge(001). The good agreement between the results from the EGM and experiments or DFT calculations show the dominant electronic confinement effect in these systems. This feature is in contrast to other systems, e.g., the formation of isolated monatomic Ga chains on Si(100) surface was attributed to the strong lateral repulsion between neighboring chains [42], and the self-assembly of dimerized monatomic Co chains on a vicinal Cu(111) surface was determined to be induced by low-temperature spin-exchange [43].

## Acknowledgements

This work was supported by National Science Foundation Grants CHE-1111500 and CHE-1404503. It was performed at Ames Laboratory, which is operated for the US Department of Energy by Iowa State University under Contract No. DE-AC02-07CH11358. The author thanks James W. Evans for a critical reading of the manuscript.

## Appendix

In 2006, Chiu *et al.* found a series of magic numbers in single-atomic-layer thick Ag nanoislands (nanopucks) grown on Pb(111) islands from their STM experiments [39]. The Ag nanopucks with *N* = 6, 7, 8, 10, 12, ... atoms are preferably observed. To explain these magic numbers, two effects should be taken into account: one is the geometrical shell-closing effect, and another one is the electronic confinement effect. The geometrical effect can be analyzed by 2D close packing of Ag atoms [39]. Here let us examine the electronic confinement effect by using the EGM to the Ag nanopucks.

For simplicity, these Ag nanopucks are modeled as the single-atomic-layer nanocylinders with a thickness of 1.45 nm from ionic radius of Ag atom [27]. Thus, the formulation in Sec. 2.2 can be directly applied to this system, where the length $l$ should be understood as thickness, see the inset in Fig. 8(a). Here only freestanding Ag nanocylinders are calculated, and therefore no charge spillage related to substrate is included, so that the results from the EGM can be compared with the available data from Chiu *et al.*'s DFT calculations for the freestanding Ag nanopucks [39]. Ag has one 5s valence electron, and then the number, *N*, of atoms in a freestanding Ag nanocylinder is the same as its number of electrons.

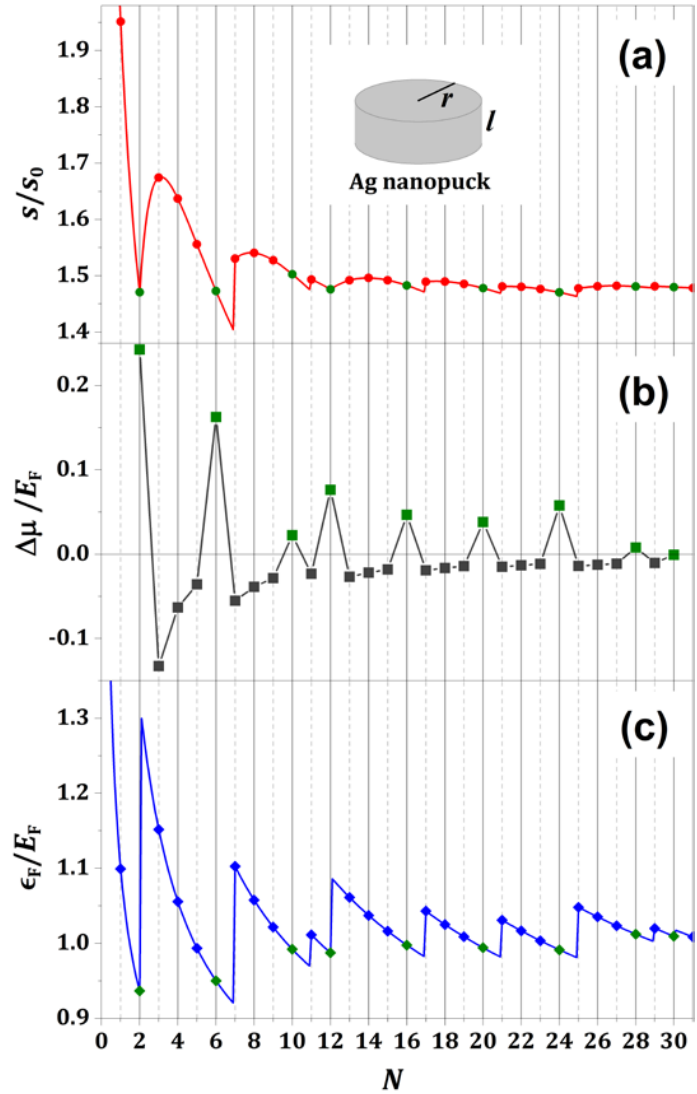

**Figure 8** (a) Surface free energy *s* from Eq. (10) versus the number *N* of atoms of single-atomic-layer Ag nanocylinders. The green dots correspond to magic *N*. Inset: The shadowed cylinder represents the geometry of uniform positive-charge background of an Ag nanopuck. (b) The stability index $\Delta\mu_N$ versus *N*. (c) Fermi energy $\epsilon_F$ from Eq. (9) versus *N*.

Figure 8(a) shows the curve of surface free energy $s$ from Eq. (10) versus $N$ (and therefore radius $r$), and Fig. 8(b) plots the total-energy second difference $\Delta\mu_N = E_{N+1} + E_{N-1} - 2E_N$, which is a stability index: the cluster with $\Delta\mu_N \geq 0$ is stable [1]. Thus, from Fig. 8(b), the magic numbers are $N$ = 2, 6, 10, 12, 16, 20, 24, 28, 30 …, which is in overall agreement with Chiu *et al.*'s DFT results: $N$ = 6, 8, 10, 12, 16, 22, 24, … [39]. The inconsistency of some magic numbers in the above two lists is quite expectable because the geometrical close-packing effect [39] is not considered in the EGM calculations. Figure 8(c) shows the Fermi energy $\epsilon_F$ from Eq. (9) versus $N$. From Fig. 8(c), it is clear that the magic numbers from the EGM correspond to the fully-occupied Fermi energy levels with the twofold or fourfold degeneracy.

In the above calculations, the Fermi wavelength $\lambda_F$ is taken to be ~ 0.633 nm, which corresponds to an fcc structure of Ag with bulk Pb lattice constant $a_{Pb} = 0.495$ nm by assuming Ag atoms ideally matches the lattice sites of substrate Pb(111) surface. More EGM results show that the curve shapes in Fig. 8 are very insensitive to the choice of Fermi wavelength $\lambda_F$ provided it is chosen to be not too small. In addition, because the thickness of the nanocylinder is small enough (only single-atomic-layer thick), the axial quantum number $n$ of occupied states are actually always the constant integer of 1. As $r$ increases, electrons only occupy the quantum states with radial quantum numbers $m$ and α taking larger integers, and then the 3D potential well is reduced to be 2D except for a constant term related to $n = 1$, in contrast to 1D potential well for the Ir nanowire or Au chain discussed in Sec. 2.